# A monte carlo density functional theory for the competition between inter and intramolecular association in inhomogeneous fluids

Bennett D. Chapman[a,1], Alejandro J. García-Cuéllar[b] and Walter G. Chapman[a]

[a]Department of Chemical and Biomolecular Engineering, Rice University, 6100 S. Main, Houston, Texas 77005

[b]Department of Mechanical Engineering, Tecnológico de Monterrey, Av. Eugenio Garza Sada 2501, Monterrey, N.L. 64849, México

## Abstract

A monte carlo density functional theory is developed for chain molecules which both intra and intermolecularly associate. The approach can be applied over a range of chain lengths. The theory is validated for the case of an associating 4-mer fluid in a planar hard slit pore. Once validated the new theory is used to study the effect of chain length and temperature on the competition between intra and intermolecular association near a hard wall. We show that this competition enhances intramolecular association near wall contact and inverts the chain length dependence of the fraction bonded intermolecularly in the inhomogeneous region.

## Keywords

---

[1] Author to whom correspondence should be addressed
Email: bennettd1980@gmail.com

# I: Introduction

Hydrogen bonding plays a crucial role in the behavior of fluids in both the physical and biological sciences. Hydrogen bonding is responsible for the remarkable properties of water[1] as well as the precise conformations of folded proteins[2] to which we owe our very existence. In the last few decades researchers began employing the hydrogen bond to do bottoms up self assembly of polymers and other nanomaterials into predetermined structures creating a new generation of smart temperature responsive materials.[3] Hydrogen bonding (or association in general) in polymer systems has been used to develop novel drug delivery methods[4], re-entrant phase behavior and temperature dependant transitions in electrical conductivity[5], form reversible cross links in rubbers[6] and create supramoleclular structures such as comb-shaped polymer assemblies[7].

Theoretical methods to describe inhomogeneous associating polymers include the random phase approximation[8], self consistent field theory[9] and classical density functional theory[10, 11]. A particularly interesting application of these methods is in the phase behavior of mixtures of *A* homopolymer and *B* homopolymer (each having a single association site) allowing for the reversible supramolecular formation of diblock copolymers which show re-entrant phase behavior between disordered homogeneous phases and bulk macrophase separation as well as re-entrant phase behavior involving lamellar microphase separation.[9, 11]

Of the previously mentioned theoretical methods, none can account for the possibility of intramolecular association. However, there are many instances in nature where intramolecular association in chain molecules plays a crucial role, such as the phase behavior telechelic polymers[12, 13] and the folding of proteins[14]. In the previously mentioned studies of the self assembly of supramolecular block copolymers[8, 9, 11], each homopolymer only had a single association site; however, if one were to consider short homopolymers with association sites on each end, intramolecular association could play a key role in the phase behavior. The balance between intermolecular association, intramolecular association and Van der Waals forces could result in novel phase behavior. Also, it is well known that glycol ethers exhibit a significant degree of intramolecular hydrogen bonding.[15] Due to the low molecular weight and high boiling point glycol ethers hold promise as green solvents; a theory for inhomogeneous chain molecule fluids which both intra and intermolecularly associates is needed to study the interfacial behavior of these fluids.

The bulk phase behavior of inter and intramolecularly associating chain fluids was a problem tackled in the 1990's by Sear and Jackson[16] (SJ) and Ghonasgi and Chapman[17, 18]. SJ developed a bulk equation of state in the framework of Wertheim's thermodynamic perturbation theory TPT[19-23]. Wertheim developed a multi density statistical mechanical formalism, where each bonding state of a molecule is assigned a density, which is ideal for short range directional interactions. By design, Wertheim's theory can reproduce the saturation of hydrogen bonds through exact graphical cancelations. The theory is most often applied as a perturbation theory which treats association as a perturbation to a hard sphere reference fluid.[23] Complex polyatomic molecules can be created from a hard sphere fluid by decorating the spheres with association sites and letting the association energies become infinitely large. In first order perturbation theory[23],TPT1, only association between pairs of spheres is considered. This results in overly flexible chains in which there is no intramolecular correlations between second nearest neighbors and beyond.

Another approximation made in TPT1 is the neglect of all ring graphs which account for rings of association bonds. This was the problem specifically addressed for homogeneous systems by SJ[16] who included a ring graph in the fundamental graph sum to account for associated rings. To develop a theory for chains which can both inter and intramoleculary associate, SJ considered a mixture of $m$ species of spherical segments, each of which is decorated with an $A$ and $B$ association site. The order of association was restricted such that site $B$ on segment 1 could only bond with site $A$ on segment 2, site $B$ on segment 2 could only bond with site $A$ on segment 3 etc… until site $B$ on segment $m-1$ could only bond with site $A$ on segment $m$. The association energies of these internal sites were then allowed to become infinitely large creating a chain of hard spheres with a single $A$ association site on segment 1 and $B$ association site on segment $m$.  With the inclusion of the ring graph, this chain is allowed to both inter and intramoleculary associate.

Beyond bulk systems Wertheim's theory has found wide application for associating fluids in interfacial systems in the form of density functional theories (DFT). [10, 24-27] In DFT a grand potential functional is constructed and minimized with respect to segment densities to obtain equations for the spatially varying densities. DFT has proven to be a powerful tool in the study of interfacial systems[28-30]; a small set of applications of DFT include: adsorption of chain molecules in slit pores[31], effect of association on the polymer phase diagram[11], phase behavior of polymer - colloid mixtures[32] and orientations of rod coil molecules adsorbed at liquid interfaces[33].

Recently, Marshall et al.[34] extended the approach of SJ to interfacial systems in the context of DFT by treating Wertheim's theory in inhomogeneous form. The theory was found to be in good agreement with simulation data for the case of a fluid of 4-mers with association sites on the first and last segment in a planar slit pore. For the associating 4-mer, the competition between intra and intermolecular association results in interesting behavior at low density where it was shown that interfacial tension went through a maximum as temperature was decreased.[34] Also, it was shown that neglecting intramolecular association resulted in a significant underprediction of association in the system (as compared to simulation where the chains could intramolecularly associate). Unfortunately, the theory is very computationally demanding and can only be applied to very short chain molecules. The computation time of the theory is the result of the irreducible ring integral (the integral cannot be factored into pair contributions in the form of recursion relations). In addition, the best results for the theory are obtained if both the ring and chain are treated as self avoiding (no intramolecular overlaps). What is needed is a computationally efficient method to evaluate both the ring and chain integrals in a self avoiding fashion. This method is found in the form of monte carlo density functional theory (MCDFT).

In MCDFT the chain integrals are rewritten as an ensemble average of the external potential and density dependant terms over a single chain probability distribution function.[35-37] These ensemble averages are then evaluated using single chain monte carlo simulations which allows one to solve the ideal chain problem exactly, while avoiding computationally demanding and time consuming numerical integrals. MCDFT has been applied to DFT's[35, 36, 38] for chain molecules based on Wertheim's theory and has been found to be accurate and convenient.

Our goal in this work is to develop a new formalism to study chain molecules which both intra and intermolecularly associate. It would be numerically impractical to apply the theory of Marshall et *al.*[34] to chain molecules with more than 5 segments; for this reason we convert this DFT for intra / inter molecular associating chain molecules into the form of a MCDFT. This is a two step process, first the theory must be converted into a form based around a molecular density, and the converted theory must be written in the form of a MCDFT. The MCDFT form of the theory will require single chain and single "ring" simulations to be performed. The theory is validated against many chain simulations for a fluid of associating 4-mers in a planar slit pore. Once validated the theory is used to study the effect of chain length and temperature on the competition between intra and intermolecular association in a chain fluid confined in a hard slit pore. Specifically we wish to study how the presence of a hard wall perturbs the bulk distribution of chains associated intra or intermolecularly. This type of system can provide insight in into the behavior of glycol ethers and telechelic polymers near solid surfaces. In this paper we focus on chains with association sites on the first and last segments; however, the general approach developed here can be extended to other geometries (for instance, ten segment chains with association sites on the first and sixth segments).

## II: Theory

In this section we define the molecular model and extend the DFT of Marshall et *al.*[34] for intramolecularly associating chain molecules into a MCDFT form. In this work we follow Ghonasgi and Chapman[17, 18] and consider molecules which consist of a chain of $m$ hard spherical segments of diameter σ bonded at hard sphere contact. Each position on the chain is occupied by a certain species of segment, so there are $m$ species of segments in total. A diagram of this molecule can be found in Fig. 1. The first segment in the chain (species 1) has an association site labeled $A$ with orientation vector $\vec{r}_A$ and the last segment in the chain (segment $m$) has an association site labeled $B$ with orientation vector $\vec{r}_B$. The orientation vectors are always at an angle of $90^o$ to the vector connecting the segment with the association site to the next segment along the chain. The interaction potential between the segment types is given by

$$\phi^{(i,j)}(12) = \phi_{HS}(r_{12}) + \phi_{AB}^{(i,j)}(12) + \phi_{BA}^{(i,j)}(12) \tag{1}$$

where all segments interact with a reference hard sphere potential $\phi_{HS}(r_{12})$ and an orientation dependant association potential $\phi_{AB}^{(i,j)}(12)$ where $1 = \{\vec{r}_1, \Omega_1\}$ represents the position $\vec{r}_1$ and orientation $\Omega_1$ of segment 1. The association potential is that of a conical site[39] and is given by

$$\phi_{AB}^{(i,j)}(12) = \begin{cases} -\varepsilon_{AB}^{(i,j)}, & r_{12} \le r_c^{(i,j)}; \alpha_A \le \alpha_c; \alpha_B \le \alpha_c \\ 0 & otherwise \end{cases} \tag{2}$$

For all bonds internal to the chain $\varepsilon_{AB}^{(i,j)} \rightarrow \infty$ and $r_c^{(i,j)} = \sigma$, while the association energy between site $A$ on segment 1 and site $B$ on segment $m, \varepsilon_{AB}^{(1,m)} \rightarrow \varepsilon_{AB}$, remains finite and adjustable.

The potential in Eq. (2) states that if segment type 1 and $m$ are within a distance $r_c^{(1,m)} = r_c$ of each other (either the same chain or a different chain) and each segment is oriented such that the angle between the site orientation vector and the vector connecting the two segments $\alpha_A$ is less than some critical angle $\alpha_c$, the two sites are considered bonded and the energy of the system is decreased by a factor $\varepsilon_{AB}$. We only allow association between site $A$ on segment 1 and site $B$ on segment $m$ (in addition to chain forming bonds), that is $\varepsilon_{AA} = \varepsilon_{BB} = 0$. For this work we choose $r_c = 1.1\sigma$ and $\alpha_c = 27°$. This choice restricts association to only one bond per association site.[39]

The free energy functional of the system is given as the sum of the ideal free energy of a mixture of hard spheres $A^{ID}[\{\rho^{(k)}\}]$ where $\rho^{(k)}$ is the density of species $k$, excess contribution due to hard sphere repulsions $A^{HS}[\{\rho^{(k)}\}]$, which is modeled using Rosenfeld's fundamental measure theory[40, 41], and excess contribution due to chain formation and inter/intra molecular association $A^{WE}[\{\rho^{(k)}\}]$

$$A[\{\rho^{(k)}\}] = A^{ID}[\{\rho^{(k)}\}] + A^{WE}[\{\rho^{(k)}\}] + A^{HS}[\{\rho^{(k)}\}] \tag{3}$$

$A^{WE}[\{\rho^{(k)}\}]$ is derived in the framework of Wertheim's perturbation theory[19-23] using the ring graph of Sear and Jackson[16] and is given as[34]

$$\tag{4}$$

$$\frac{A^{WE}[\{\rho^{(k)}\}]}{k_B T} = \sum_{k=1}^{m} \int d\vec{r} \left( \rho^{(k)}(\vec{r}) \ln\left( \frac{\tilde{\rho}_o^{(k)}(\vec{r})}{\rho^{(k)}(\vec{r})} \right) + \rho^{(k)}(\vec{r}) \right) - \int d\vec{r} \rho^{(1)}(\vec{r}) \left( X_A(\vec{r}) + \chi_{ring}(\vec{r}) \right)$$

Equation 4 has been rearranged from the original reference[34] and the complete formation of the chain has been enforced. We have followed Kierlik and Rosinberg[42] and introduced scaled monomer densities $\tilde{\rho}_o^{(k)}$. These are monomer densities $\rho_o^{(k)}$, (density of spheres not bonded which go to zero in the limit of chain formation) scaled by the infinitely large magnitude of the chain forming Mayer functions $\lambda = f_{AB}\big|_{\varepsilon_{AB}\to\infty}$ where $f_{AB} = \exp(\varepsilon_{AB}/k_B T)-1$. The scaled monomer densities are given in ref [42] as $\tilde{\rho}_o^{(k)} = \lambda^{1/2}\rho_o^{(k)}$ for $k=1$ and $m$ and $\tilde{\rho}_o^{(k)} = \lambda\rho_o^{(k)}$ for $k=2$ through $m$-1 and are of order $\tilde{\rho}_o^{(k)} \sim 1$. This rescaling has no effect on the final calculated quantities (density profiles etc...). The term $X_A(\vec{r})$ is the fraction of end segments *not* bonded and $\chi_{ring}(\vec{r})$ is the fraction of end segments *bonded* intramolecularly into rings.

In this work we introduce the *m* point molecular densities, see also ref[42], which are generated from functional derivates of the functional $\Delta\tilde{c}^{(o)}$ listed in our first paper (Equation 29 in ref[34]). The result is

$$\rho^{(1...m)}(\vec{r}_1...\vec{r}_m) = \frac{\delta^m \Delta\tilde{c}^{(o)}}{\delta\tilde{\rho}_o^{(m)}(\vec{r}_m)\cdots\delta\tilde{\rho}_o^{(1)}(\vec{r}_1)} = \rho_{chains}^{(1...m)}(\vec{r}_1...\vec{r}_m) + \rho_{ring}^{(1...m)}(\vec{r}_1...\vec{r}_m) \tag{5}$$

Where the chains and ring molecular densities are given by

$$\rho_{chains}^{(1...m)}(\vec{r}_1...\vec{r}_m) = \frac{\rho_{BASE}^{(1...m)}(\vec{r}_1...\vec{r}_m)}{X_A^{poly}(\vec{r}_1)X_A^{poly}(\vec{r}_m)} \tag{6}$$

$$\rho_{ring}^{(1...m)}(\vec{r}_1...\vec{r}_m) = \frac{1}{\Omega_1\Omega_m}\int d\Omega_1 d\Omega_m \rho_{BASE}^{(1...m)}(\vec{r}_1...\vec{r}_m)y_c(\vec{r}_1,\vec{r}_m)e_R(r_{1m})\frac{\sigma^2}{r_{1m}^2}f_{AB}(1,m) \tag{7}$$

Where we note that in the integral $\int d\Omega_1 d\Omega_m$, the available orientations for the first and last segment are constrained by the set location of the association sites on the segments (see Fig. 1); for this reason the orientational integration cannot be decoupled from position in the chain term. The association Mayer function is $f_{AB}(12) = \exp\left(-\phi_{AB}^{(1,m)}(12)/k_BT\right) - 1$. The "base" molecular density $\rho_{BASE}^{(1...m)}(\vec{r}_1...\vec{r}_m)$ is of the form of a non-associating chain and is given by[42]

$$\rho_{BASE}^{(1...m)}(\vec{r}_1...\vec{r}_m) = \left(\prod_{k=1}^{m} \tilde{\rho}_o^{(k)}(\vec{r}_k)\right)\left(\prod_{k=1}^{m-1} y_c(\vec{r}_k, \vec{r}_{k+1})\right) P_{chain}(\vec{r}_1...\vec{r}_m) \qquad (8)$$

In Eq. (8) the functions $y_c(\vec{r}_k, \vec{r}_{k+1})$ are the inhomogeneous cavity correlation functions of the hard sphere reference system at hard sphere contact. The term $P_{chain}(\vec{r}_1...\vec{r}_m)$ is the probability that an isolated chain has a conformation $\{\vec{r}_1...\vec{r}_m\}$. In the original publication[34] the theory was derived for a freely jointed chain where non-adjacent segments along the chain could overlap; however, without loss of generality[43] we can restrict the chain to be self avoiding. For the remainder of this work $P_{chain}(\vec{r}_1...\vec{r}_m) \sim \prod_{k=1}^{m-1} \delta(r_{k,k+1} - \sigma) \prod_{\substack{all\ unbonded \\ pairs\ \{i,j\}}} e_R(r_{ij})$ is that of a self avoiding chain. In Eq. (6) the fractions $X_A^{poly}$ are the fraction of end segments not bonded *IF* there were no intramolecular association

$$X_A^{poly}(\vec{r}_1) = \cfrac{1}{1 + \int \rho^{(1)}(\vec{r}_2) X_A(\vec{r}_2) \cfrac{\sigma^2}{r_{12}^2} y_c(\vec{r}_1, \vec{r}_2) e_R(r_{12}) \cfrac{\int f_{AB}(12)d\Omega_1 d\Omega_2}{\Omega_1 \Omega_2} d\vec{r}_2} \qquad (9)$$

The reference system bonds $e_R(r_{12})$ are given by

$$e_R(r_{12}) = \begin{cases} 0 & for \quad r_{12} < \sigma \\ 1 & otherwise \end{cases} \tag{10}$$

In both Eqns. (7) and (9) we employed the approximation[39] that within the bond volume $r^2 y(r) \approx \sigma^2 y(\sigma)$ from which we obtain the approximate relation

$$y(\vec{r}_1, \vec{r}_2) = \frac{r_{12}^2}{r_{12}^2} y(\vec{r}_1, \vec{r}_2) \approx \frac{\sigma^2}{r_{12}^2} y_c(\vec{r}_1, \vec{r}_2) \tag{11}$$

The singlet densities are obtained from the molecular densities by[31]

$$\rho^{(k)}(\vec{r}) = \int d\vec{r}_1...d\vec{r}_m \delta(\vec{r} - \vec{r}_k) \rho^{(1...m)}(\vec{r}_1...\vec{r}_m) = \rho_{chains}^{(k)}(\vec{r}) + \rho_{ring}^{(k)}(\vec{r}) \tag{12}$$

where $\delta(x)$ is the Dirac delta function. The monomer densities are obtained through minimization of the grand free energy functional

$$\Omega[\{\rho^{(k)}\}] = A[\{\rho^{(k)}\}] - \sum_{k=1}^{m} \int d\vec{r}\,' \rho^{(k)}(\vec{r}\,') \left( \mu^{(k)} - V_{ext}(\vec{r}\,') \right) \tag{13}$$

where $\mu^{(k)}$ is the chemical potential of segment $k$ and $V_{ext}$ is the external field. The monomer densities are found to be[34]

$$\tilde{\rho}_o^{(k)}(\vec{r}_k) = \exp\left[ \lambda^{(k)}(\vec{r}_k) + \frac{\mu^{(k)}}{k_B T} \right] \tag{14}$$

Where

$$\lambda^{(k)}(\vec{r}_k) = \Xi^{(k)} - \frac{\delta A^{HS}/k_BT}{\delta\rho^{(k)}(\vec{r}_k)} - \frac{V_{ext}(\vec{r}_k)}{k_BT} \qquad (15)$$

And the term $\Xi^{(k)}$ is given by

$$(16)$$

$$\Xi^{(k)} = \sum_{\beta=1}^{m-1} \int \rho^{(\beta,\beta+1)}(\vec{r}_1,\vec{r}_2) \frac{\delta \ln y_c^{(\beta,\beta+1)}(\vec{r}_1,\vec{r}_2)}{\delta\rho^{(k)}(\vec{r}_k)} d\vec{r}_1 d\vec{r}_2 + \int \rho_{ring}^{(1,m)}(\vec{r}_1,\vec{r}_m) \frac{\delta \ln y_c(\vec{r}_1,\vec{r}_m)}{\delta\rho^{(k)}(\vec{r}_k)} d\vec{r}_1 d\vec{r}_m$$

$$+ \int \rho^{(1)}(\vec{r}_1)\rho^{(m)}(\vec{r}_2) X_A(\vec{r}_1) X_A(\vec{r}_2) y_c(\vec{r}_1,\vec{r}_2) \frac{\sigma^2}{r_{12}^2} \frac{\int f_{AB}(12) d\Omega_1 d\Omega_2}{\Omega_1 \Omega_2} e_R(r_{12}) \frac{\delta \ln y_c(\vec{r}_1,\vec{r}_2)}{\delta\rho^{(k)}(\vec{r}_k)} d\vec{r}_1 d\vec{r}_2$$

The two point molecular densities are obtained from the relation[31]

$$\rho^{(k,j)}(\vec{r},\vec{r}') = \int d\vec{r}_1...d\vec{r}_m \delta(\vec{r}-\vec{r}_k)\delta(\vec{r}'-\vec{r}_j)\rho^{(1...m)}(\vec{r}_1...\vec{r}_m) \qquad (17)$$

There is a similar relationship between $\rho_{ring}^{(1,m)}(\vec{r}_1,\vec{r}_m)$ and the ring molecular density Eq. (7). Now we eliminate all of the monomer densities in $\rho_{BASE}^{(1...m)}(\vec{r}_1...\vec{r}_m)$ to obtain

$$\rho_{BASE}^{(1...m)}(\vec{r}_1...\vec{r}_m) = \exp\left[\frac{\mu_M}{k_BT} + \sum_{k=1}^{m} \lambda^{(k)}(\vec{r}_k) + \sum_{k=1}^{m-1} \ln y_c(\vec{r}_k,\vec{r}_{k+1})\right] P_{chain}(\vec{r}_1...\vec{r}_m) \qquad (18)$$

Where $\mu_M = \sum_{k=1}^{m} \mu_k$ is the total chemical potential and is given by[34]

$$\frac{\mu_M}{k_BT} = \frac{\mu_{HS}}{k_BT} + \ln X_A + \ln X_A^{poly} - (m-1)\ln y_c - (m-1)\eta\frac{\partial \ln y_c}{\partial \eta}$$

$$- (1-X_A)\eta\frac{\partial \ln y_c}{\partial \eta} - (m-1)\ln\rho \qquad (19)$$

Where $\mu_{HS}$ is the chemical potential of the hard sphere reference system and $\eta = \frac{\pi}{6}m\rho\sigma^3$ is the packing fraction. Three equations govern the various bonding fractions: the first is Eq. (9) and the remaining two are given below[34]

$$X_A(\vec{r}) = X_A^{poly}(\vec{r})\left(1 - \chi_{ring}(\vec{r})\right) \tag{20}$$

Where $\chi_{ring}(\vec{r})$ is determined self consistently through the relation

$$\chi_{ring}(\vec{r}) = \frac{\rho_{ring}^{(1)}(\vec{r})}{\rho_{ring}^{(1)}(\vec{r}) + \rho_{chains}^{(1)}(\vec{r})} \tag{21}$$

The form of the theory is significantly simplified if we consider the following approximation of $y_c(\vec{r}_1, \vec{r}_2)$. In the free energy Eq. (4) the contact cavity correlation functions are embedded in integrals of the form $\int y_c(\vec{r}_1, \vec{r}_2)\delta(r_{12} - \sigma)S(\vec{r}_1, \vec{r}_2)d\vec{r}_1 d\vec{r}_2$ where $S$ is some function and $\delta(x)$ is the Dirac delta function. It is reasonable to assume that when the location of a given particle in the integral is fixed and the other particle is integrated over the surface that a coarse grained approximation of $y_c(\vec{r}_1, \vec{r}_2)$ can be used. To this end we consider the averaged cavity correlation function $\overline{y_c(\vec{r}_1)}$

$$\overline{y_c(\vec{r}_1)} = \frac{\int d\vec{r}_2\, y_c(\vec{r}_1, \vec{r}_2)\delta(r_{12} - \sigma)}{4\pi\sigma^2} \tag{22}$$

Where we have averaged the pair function on the surface of a sphere of diameter $\sigma$ surrounding $\vec{r}_1$. Since we could have also averaged over $\vec{r}_1$ to obtain $\overline{y_c(\vec{r}_2)}$ we average these two results using a geometric mean

$$\ln y_c(\vec{r}_1, \vec{r}_2) = \frac{1}{2} \ln \left( \overline{y_c(\vec{r}_1)} \times \overline{y_c(\vec{r}_2)} \right) \qquad (23)$$

Now we assume the coarse grained pair function can be evaluated as the bulk contact cavity correlation function $y = (1 - 0.5\eta)/(1 - \eta)^3$ evaluated with a coarse grained density, that is

$$\overline{y_c(\vec{r}_1)} \approx y_c(\overline{\eta}(\vec{r}_1)) \qquad (24)$$

Where $\overline{\eta}(\vec{r}_1) = \frac{\pi}{6} \sigma^3 \sum_{k=1}^{m} \overline{\rho}^{(k)}(\vec{r}_1)$ and $\overline{\rho}^{(k)}(\vec{r}_1)$ is some coarse grained density. Here we choose the average to be taken in a spherical volume of radius $\sigma$

$$\overline{\rho}^{(k)}(\vec{r}_1) = \frac{3}{4\pi\sigma^3} \int d\vec{r}_2 \rho^{(k)}(\vec{r}_2) H(\sigma - r_{12}) \qquad (25)$$

where $H(x)$ is the Heaviside step function. Equations (22) – (25) give the approximation of the pair cavity correlation function. This approximation has been show to give accurate results in inhomogeneous systems.[44-46] With this approximation of the cavity functions we can simplify Eq. (16) to

$$\begin{aligned}
\Xi^{(k)} = \frac{1}{2} \sum_{\beta=1}^{m-1} \int \left( \rho^{(\beta)}(\vec{r}_1) + \rho^{(\beta+1)}(\vec{r}_1) \right) \frac{\delta \ln y_c(\overline{\eta}(\vec{r}_1))}{\delta \rho^{(k)}(\vec{r}_k)} d\vec{r}_1 \\
+ \int \rho^{(1)}(\vec{r}_1) \left( 1 - X_A(\vec{r}_1) \right) \frac{\delta \ln y_c(\overline{\eta}(\vec{r}_1))}{\delta \rho^{(k)}(\vec{r}_k)} d\vec{r}_1
\end{aligned} \qquad (26)$$

Equation (26) completes the density functional theory.

What we have done is introduce molecular densities into the framework of our previous theory;[34] now we show how the singlet densities can be written as ensemble averages over single molecule distribution functions. We begin with the singlet chains density given by

$$\rho_{chains}^{(k)}(\vec{r}) = \int d\vec{r}_1...d\vec{r}_m \delta(\vec{r}-\vec{r}_k)\rho_{chains}^{(1...m)}(\vec{r}_1...\vec{r}_m) = \int d\vec{r}_1...d\vec{r}_m \delta(\vec{r}-\vec{r}_k)\frac{\rho_{BASE}^{(1...m)}(\vec{r}_1...\vec{r}_m)}{X_A^{poly}(\vec{r}_1)X_A^{poly}(\vec{r}_m)} \tag{27}$$

Now using Eq. (18) and rearranging

$$\rho_{chains}^{(k)}(\vec{r}) = \exp\left(\frac{\mu_M}{k_BT}\right)\int d\vec{r}_1...d\vec{r}_m \delta(\vec{r}-\vec{r}_k)P_{chain}(\vec{r}_1...\vec{r}_m)$$
$$\times \exp\left[\sum_{j=1}^{m}\lambda^{(j)}(\vec{r}_j) + \sum_{j=1}^{m-1}\ln y_c(\vec{r}_j,\vec{r}_{j+1}) - \ln X_A^{poly}(\vec{r}_1) - \ln X_A^{poly}(\vec{r}_m)\right] \tag{28}$$

Equation (28) can now be written as an ensemble average over the single chain distribution function $P_{chain}(\vec{r}_1...\vec{r}_m)$[37]

$$\rho_{chains}^{(k)}(\vec{r}) = \exp\left(\frac{\mu_M}{k_BT}\right)\left\langle \exp\left[\sum_{j=1}^{m}\lambda^{(j)}(\vec{r}_j) + \sum_{j=1}^{m-1}\ln y_c(\vec{r}_j,\vec{r}_{j+1}) - \ln X_A^{poly}(\vec{r}_1) - \ln X_A^{poly}(\vec{r}_m)\right]\right\rangle_{P_{chain}}^{\vec{r}=\vec{r}_k} \tag{29}$$

The bracket $\left\langle\ \right\rangle_{P_{chain}}^{\vec{r}=\vec{r}_k}$ represents the conformational average over the distribution function $P_{chain}$ with segment $k$ fixed at point $\vec{r}$ in the fluid. This ensemble average is evaluated over all chain conformations using a single chain monte carlo simulation. We can rewrite the ring density as

$$\rho_{ring}^{(1...m)}(\vec{r}_1...\vec{r}_m) = \frac{f_{AB}}{\Omega_1\Omega_m}\int d\Omega_1 d\Omega_m \rho_{BASE}^{(1...m)}(\vec{r}_1...\vec{r}_m)y_c(\vec{r}_1,\vec{r}_m)\xi(r_{1m})U(\Omega_1,\Omega_m)\frac{\sigma^2}{r_{1m}^2} \tag{30}$$

Where the function $U(\Omega_1, \Omega_m)$ is given by

$$U(\Omega_1, \Omega_m) = \begin{cases} 1 & for\ \alpha_A \le \alpha_c\ and\ \alpha_B \le \alpha_c \\ 0 & otherwise \end{cases} \tag{31}$$

and the function $\xi(r_{1m})$ is given by

$$\xi(r_{1m}) = \begin{cases} 1 & \sigma \le r_{1m} \le r_c \\ 0 & otherwise \end{cases} \tag{32}$$

Using Eq. (18) the ring singlet density is now obtained as

$$\rho_{ring}^{(k)}(\vec{r}) = \int d\vec{r}_1 ... d\vec{r}_m \delta(\vec{r} - \vec{r}_k) \rho_{ring}^{(1...m)}(\vec{r}_1 ... \vec{r}_m)$$

$$= f_{AB} \exp\left(\frac{\mu_M}{k_B T}\right) \int d\vec{r}_1 ... d\vec{r}_m d\Omega_1 d\Omega_m \delta(\vec{r} - \vec{r}_k) \xi(r_{1m}) \frac{U(\Omega_1, \Omega_m)}{\Omega_1 \Omega_m} P_{chain}(\vec{r}_1 ... \vec{r}_m)$$

$$\times \exp\left[\sum_{j=1}^{m} \lambda^{(j)}(\vec{r}_j) + \sum_{j=1}^{m-1} \ln y_c(\vec{r}_j, \vec{r}_{j+1}) + \ln y_c(\vec{r}_1, \vec{r}_m) + 2\ln \frac{\sigma}{r_{1m}}\right] \tag{33}$$

There are 2 methods to evaluate Eq. (33) as an ensemble average. The first is to write the ring integral as an ensemble average over the chain distribution function $P_{chain}(\vec{r}_1 ... \vec{r}_m)$ giving

$$\rho_{ring}^{(k)}(\vec{r}) = f_{AB} \exp\left(\frac{\mu_M}{k_B T}\right) \left\langle \int d\Omega_1 d\Omega_m \frac{U(\Omega_1, \Omega_m)}{\Omega_1 \Omega_m} \xi(r_{1m}) \right.$$

$$\left. \times \exp\left[\sum_{j=1}^{m} \lambda^{(j)}(\vec{r}_j) + \sum_{j=1}^{m-1} \ln y_c(\vec{r}_j, \vec{r}_{j+1}) + \ln y_c(\vec{r}_1, \vec{r}_m) + 2\ln \frac{\sigma}{r_{1m}}\right] \right\rangle_{P_{chain}}^{\vec{r} = \vec{r}_k} \tag{34}$$

Equation (34) is the direct method to evaluate the ensemble average of the ring integral, the average is over the conformations of a single chain. Unfortunately, the vast majority of chain conformations are not valid ring states, so the evaluation of Eq. (34) through single chain monte carlo simulation is very inefficient. Alternatively, we can assume a ring is already formed and define a ring distribution function

$$P_{ring}(\vec{r}_1...\vec{r}_m) \sim P_{chain}(\vec{r}_1...\vec{r}_m)\xi(r_{1m}) \qquad (35)$$

Now, if we wrote the single ring density as an ensemble average over the ring distribution function $P_{ring}(\vec{r}_1...\vec{r}_m)$ renormalized to the number of ring states, the ring density would be overestimated due the fact that at no point did we account for the fact that a ring is conformationally constrained as compared to chain. There are much fewer available ring states than chain states. To correct for this we introduce a probability $P_{form}(m)$ which gives the probability that a chain of length $m$ has the correct conformation and segments 1 and $m$ have the correct orientation that a ring can be formed. Since the single "ring" simulation is independent of the external field and density profiles we shall assume that this probability to be that of an isolated chain, independent of density and given by the ratio of ring states to chain states

$$P_{form}(m) = \frac{\int d\Omega_1 d\Omega_m d\vec{r}_1...d\vec{r}_m U(\Omega_1,\Omega_m)P_{chain}(\vec{r}_1...\vec{r}_m)\xi(r_{1m})}{\Omega_1\Omega_m\int d\vec{r}_1...d\vec{r}_m P_{chain}(\vec{r}_1...\vec{r}_m)} \qquad (36)$$

Equation (36) is generally valid for a chain with association sites on each end segment. Specifically for the model described in Fig. 1 (where the site orientation vectors $\vec{r}_A$ and $\vec{r}_B$ are restricted to be at a $90°$ angle to the vectors $\vec{r}_{2,1} = \vec{r}_2 - \vec{r}_1$ and $\vec{r}_{m-1,m}$ respectively) there are only $\Omega = 2\pi$ independent orientations for each of the end segments and the orientation vectors are restricted to be on the edge of a circle whose plane is normal to the vectors $\vec{r}_{2,1}$ or $\vec{r}_{m-1,m}$. With these restrictions the orientational integration of the end segments is $\int d\Omega \rightarrow \int_{0}^{2\pi} d\gamma$, where $\gamma$ is an angle in the plane normal to the vector $\vec{r}_{2,1}$ for site A and $\vec{r}_{m-1,m}$ for site B. For this model Eq. (36) was evaluated using monte carlo integration for a number of chain lengths, the results can be found in table 1. Now the ring singlet densities are given by the monte carlo ensemble average over the normalized probability distribution $P_{ring}(\vec{r}_1...\vec{r}_m)$ of the following

$$(37)$$

$$\rho_{ring}^{(k)}(\vec{r}) = f_{AB} P_{form}(m) \exp\left(\frac{\mu_M}{k_B T}\right) \left\langle \exp\left[\sum_{j=1}^{m} \lambda^{(j)}(\vec{r}_j) + \sum_{j=1}^{m-1} \ln y_c(\vec{r}_j, \vec{r}_{j+1}) + \ln y_c(\vec{r}_1, \vec{r}_m) + 2\ln\frac{\sigma}{r_{1m}}\right]\right\rangle_{P_{ring}}^{\vec{r}=\vec{r}_k}$$

The product $f_{AB} P_{form}(m)$ represents the probability of ring formation with $f_{AB}$ being the energetic contribution and $P_{form}(m)$ the entropic. Equation (37) completes the development of the MCDFT for associating chain molecules with one site located on each end of the chain.

In this work we will study the behavior of a fluid of associating chains in a planar slit pore with external potential

$$V_{ext}(z) = \begin{cases} \infty & for \quad z < 0 \quad or \quad z > H \\ 0 & otherwise \end{cases} \qquad (38)$$

With a 1-D inhomogeneity in the z dimension Eq. (9) can be simplified as

$$X_A^{poly}(z_1) = \frac{1}{1 + f_{AB}\dfrac{\pi}{2}(1-\cos\alpha_c)^2(r_c-\sigma)\sigma \displaystyle\int_{z_1-\sigma}^{z_1+\sigma} dz_2 \rho^{(1)}(z_2) X_A(z_2) y_c(z_1, z_2)} \qquad (39)$$

The overall methodology to obtain the singlet densities is as described in our previous paper[34]; we first specify a bulk chain density $\rho$ which allows us to calculate the bulk $X_A$ from the relation[17]

$$\left(\frac{1}{X_A}\right)^3 + \left(\rho\Delta^{\text{inter}} - \Delta^{\text{intra}} - 1\right)\left(\frac{1}{X_A}\right)^2 - 2\rho\Delta^{\text{inter}}\left(\frac{1}{X_A}\right) - \left(\rho\Delta^{\text{inter}}\right)^2 = 0 \tag{40}$$

where $\Delta^{\text{inter}} = \pi\sigma^2\left(1 - \cos\alpha_c\right)^2\left(r_c - \sigma\right)f_{AB}\,y_c$ and $\Delta^{\text{intra}}$ is obtained through the bulk limit of the MCDFT as

$$\Delta^{\text{intra}} = f_{AB}P_{form}(m)\left\langle\frac{\sigma^2}{r_{1m}^2}\right\rangle_{P_{ring}} y_c \tag{41}$$

The ensemble average $\left\langle\sigma^2/r_{1m}^2\right\rangle_{P_{ring}}$ is very accurately given by $\left\langle\sigma^2/r_{1m}^2\right\rangle_{P_{ring}} \approx \dfrac{4}{\left(r_c/\sigma + 1\right)^2}$ .

Armed with $X_A$ the fraction $X_A^{poly}$ can be calculated through the bulk limit of Eq. (9) as $X_A^{poly} = 1/\left(1 + \rho X_A \Delta^{\text{inter}}\right)$, from which we can obtain the ring fraction through Eq. (20). With all of the bulk bonding fractions calculated the bulk chemical potential is calculated through Eq. (19). With the bulk problem solved, Eqns. (12) for the segment density profiles (with the chains densities and ring densities given by Eq. (29) and (37) respectively) and Eq. (21) for $\chi_{ring}$ are solved using a Picard iteration where bulk densities and ring fractions are used as an initial guess.

## III: Single Chain and Single Ring Simulations

The progress of the single chain and single ring simulations given by Eqns. (29) and (37) are independent of the external potential and any density dependant terms, meaning that a generated conformation is used to evaluate the densities at each point in the domain. The chain conformations to evaluate Eqns. (29) for the chain integral are produced by generating azimuthal angles $\{0 \le \phi_j \le 2\pi\}$ and cosine of the polar angle $\{-1/2 \le \cos\theta_j \le 1\}$ in a coordinate system for segment $j$ centered on segment $j-1$ and whose z – axis is parallel to the bond vector $\vec{r}_{j-1,j-2} = \vec{r}_{j-1} - \vec{r}_{j-2}$. This choice of coordinate system guarantees no overlap between second nearest neighbors. If there is no overlap between segments in the chain the conformation is accepted. The acceptance rate for a total chain conformation using this approach is 100% for $m = 3$, 95 % for $m = 4$, 81% for $m = 6$. In total $\sim 10^5$ independent chain conformations were used to evaluate the chain densities.

Generating ring conformations to evaluate Eq. (37) is more difficult than the chain case due to the fact that the vector connecting the first and last sphere in the ring must satisfy the relation $\{\sigma \le r_{m1} \le r_c\}$. Due to this restriction the method used to evaluate the chain integral would be very inefficient since there is a low probability that a randomly generated chain conformation would be in a valid ring state. Instead we follow an approach similar to that of Dickman and Hall[47] who obtained chain conformations by shaking a starting conformation. Here we start with the chain in a valid ring conformation and then we subject each bond vector to an independent random displacement $\vec{r}_{j,j-1} \rightarrow \vec{r}_{j,j-1} + \delta \times \vec{b}$ where $\vec{b}$ is a random vector with components generated in the range $\{-1 \le b_k \le 1\}$ and $\delta$ is a constant which varies between $\{0 \le \delta \le 1\}$ chosen such that the acceptance rate in approximately 50%. The resulting vector is normalized such that $\left| \vec{r}_{j,j-1} \right| = 1$. After generating a new conformation if there is no segment overlap in the chain and the chain is still in a valid ring state the move is accepted, otherwise the old conformation is accepted. Averages were taken every 5 trial conformations with a total of ~ $10^6$ trial conformations generated.

The evaluation of the chain integral was much faster than that for the ring integral with the computation time increasing as chain length increases. In this work approximately the same number of conformations where used for each chain length $m = 3 - 7$. To obtain a converged density profile for a 4-mer using the MCDFT, where the bulk density is used as an initial guess, only takes a couple of hours on a Dell laptop; in contrast, the many chain monte carlo simulation would take approximately 1 - 2 days between equilibration and averaging, if it could be performed at all. In our previous work[34] many chain monte carlo simulations for 4 - mers which both intra and intermolecular associate could not be performed at $\eta = 0.3$ for $\varepsilon^* > 6$ because the bonding fractions would not converge over the entire domain. A similar problem was encountered when performing many chain simulations for a 4-mer chain at $\eta = 0.1$ for $\varepsilon^* = 10$. It should be noted that the MCDFT is in no way limited in this aspect and can be applied over the full range of $\varepsilon^*$. Many chain monte carlo simulations would become much more difficult as chain length increases due to the rapid decrease in the probability two chain ends are positioned and oriented such that intramolecular association can occur, see table 1. For instance $P_{form}(7)/P_{form}(4) \sim 10^{-1}$ which means obtaining good statistics while sampling $\chi_{ring}(z)$ would become increasingly difficult as chain length is increased, especially near wall contact. Of course, computation time of the MCDFT would also increase with increasing chain length due to the larger number of single chain conformations which would be needed. Most of this increase in computation time would be associated with the ring integral; however, for long chains the probability of intramolecular association would be small and the ring integral could be safely neglected.

## IV: Validation

In this section we validate the derived MCDFT by comparison to many chain monte carlo simulation data for the case of a 4 – mer of the type described in Fig. 1 in a planar slit pore. In this work we choose $r_c = 1.1\sigma$ and $\alpha_c = 27°$. Mainly we compare to NVT monte carlo simulations from our previous paper[34]; however when considering intramolecular association only we perform additional simulations at average pore packing fractions of $\eta_{av} = 0.3$ for reduced association energies $\varepsilon^* = \varepsilon_{AB} / k_B T = 7,8$. These simulations are performed in the same manner as described previously.[34] We also include density profiles for chain molecules which can only intramolecularly associate, these were not shown in our previous publication[34]. Since results are symmetric about the center of the pore, results are presented versus distance from one pore wall.

We begin by considering 4 – mers which can only intramolecularly associate. To neglect intermolecular association in the theory we simply set $X_A^{poly} = 1$. Figure 2 compares theoretical and simulation density profiles at packing fractions $\eta_{av} = 0.1, 0.3$ and reduced association energies $\varepsilon^* = 0, 8$. At the low $\eta_{av} = 0.1$ (which corresponds to a bulk packing fraction $\eta_b \approx 0.1$) the fluid is depleted near the wall in a drying effect typical of polyatomic fluids due to the loss of conformational entropy near the wall. Increasing the density to the liquid like $\eta_{av} = 0.3$ (which corresponds to a bulk packing fraction $\eta_b \approx 0.29$) the fluid wets the wall due to the hard sphere packing effect. At each density, increasing $\varepsilon^*$ decreases the wall contact density of the end segment and increases the contact density of the middle segment. Overall the simulation and theory are in excellent agreement. Treating the ring and chain as self avoiding gives improved results over our previous theory where the chain was treated in first order perturbation theory and non – adjacent segments along the chain could overlap. [34]

Figure 3 gives the fraction of end segments bonded intramolecularly $\chi_{ring}$ for an intramolecularly associating chain fluid over a range of association energies for average packing fractions of $\eta_{av} = 0.1, \, 0.3$. For each case $\chi_{ring}$ is depleted at wall contact and goes through a maximum near $z \sim 1.1\sigma$ at $\eta_{av} = 0.1$ and $z \sim \sigma$ at $\eta_{av} = 0.3$. The decrease in the location of the maximum at high density is due to the packing of the rings near the wall. At $\eta_{av} = 0.1$ the simulation and theory are in near perfect agreement, while at $\eta_{av} = 0.3$ the theory overpredicts the amount of association. The overprediction of $\chi_{ring}$ at $\eta_{av} = 0.3$ is a result of the fact that the density dependence of $\Delta^{intra}$ (Eq. 41) is too strong. This can be seen in Fig. 4 which compares the predictions of Eq. (41) to the many chain monte carlo simulation results for $\Delta^{intra}$ of Ghonasgi and Chapman[17]. At low density, both theory and simulation are in good agreement; however, at high density the theory predicts values of $\Delta^{intra}$ which are a little large. This discrepancy at high density can be traced back to the ring graph[16] used in the development of the original DFT[34] which contains a product of pair correlation functions such that each bonded pair shares a $g(\vec{r}_1, \vec{r}_2)$. This must be the simplification of a more general graph which contains an $n$ – body correlation function (where the ring contains $n$ spheres) $g(\vec{r}_1...\vec{r}_n)$. To obtain the ring graph of Sear and Jackson[16], which we modify here, one takes the superposition

$$g(\vec{r}_1...\vec{r}_n) = g(\vec{r}_1,\vec{r}_n)\prod_{k=1}^{n-1} g(\vec{r}_k,\vec{r}_{k+1}) \prod_{\substack{all \ unbonded \\ pairs \ \{i,j\}}} e_R(r_{ij}) \qquad (42)$$

where we have included an $e_R$ bond between each pair of unbounded spheres in the ring to make the molecule self avoiding. Equation 42 is exact in the low density limit and will become less accurate at higher densities; it is for this reason the theory is most accurate at low densities. Another possible source of error is the approximation that within the bond volume $r^2 y(r) \approx \sigma^2 y(\sigma)$; we are currently in the process of improving this approximation.

Now we turn our attention to chain molecules which both intra and intermolecularly associate. The density profiles are qualitatively similar to those shown in Fig. 2, so for brevity we do not include these. Figure 5 gives the fraction of end segments which are associated (either inter or intramolecularly) $\chi = 1 - X_A$ and the fraction of end segments intramolecularly associated into rings $\chi_{ring}$ at a packing fraction of $\eta_{av} = 0.1$. As can be seen, the theory and simulation are in excellent agreement. Figure 6 gives the fractions $\chi_{ring}$ and $\chi$ at an average packing fraction $\eta_{av} = 0.3$. The overall fraction $\chi$ is in excellent agreement with simulation, while the theory predicts values of $\chi_{ring}$ which are too high.

We have shown that the new MCDFT is accurate in comparison to monte carlo simulations for the case of a 4-mer fluid in a slit pore. At high densities the theory predicts ring fractions which are a too large, however the effects of intramolecular association are most pronounced at low densities[17, 34], so we do not consider this a serious flaw in the approach. In the following section we will use the MCDFT to study the effect of chain length and temperature on fractions $\chi_{ring}$ and $\chi_{chain} = \chi - \chi_{ring}$ (fraction of end segments bonded intermolecularly) near a hard wall.

# V: Bonding fractions near a hard wall

In this section we will use the MCDFT to study the effect of $\varepsilon^*$ and chain length $m$ on the bonding fractions $\chi_{ring}$ and $\chi_{chain} = \chi - \chi_{ring}$ at a low bulk packing fraction of $\eta_b = 0.1$. We stick to the low density case due to the fact that is at low density that the effects of intramolecular association are most pronounced[17, 34]; also, it is in this realm where the theory is most accurate. At the bulk packing fraction of $\eta_b = 0.1$ the hard chain fluid can be thought of as a semi-dilute polymer solution in a theta solvent.

Before considering the inhomogeneous case we will first consider the fractions in a bulk homogeneous system, Fig. 7. For $m = 3$ it is geometrically impossible for intramolecular association to occur for chain molecules of the type given in Fig. 1, $P_{form}(3) = 0$, so $\chi_{ring}^{bulk} = 0$ for all $\varepsilon^*$. Since there is no competition between intra and intermolecular association, $\chi_{chain}^{bulk}$ increases to a limiting value of 1 at high $\varepsilon^*$. Increasing the chain length to $m = 4$ intramolecular association becomes important. In fact, it is at this chain length that $P_{form}(m)$ attains it's largest value, meaning that intramolecular association is a maximum at $m = 4$. This can be seen in the bonding fractions in Fig. 7; initially $\chi_{chain}^{bulk}$ increases with $\varepsilon^*$; however, at $\varepsilon^* \sim 10$ $\chi_{chain}^{bulk}$ reaches a maximum and begins to decrease upon increasing $\varepsilon^*$. This maximum exist due to the fact that intramolecular association dominates at large $\varepsilon^*$, at this chain length, as evidenced by the fact that $\chi_{ring}^{bulk}$ approaches unity for large $\varepsilon^*$. Increasing the chain length to $m = 5$ and higher destroys this maximum in $\chi_{chain}^{bulk}$, due to the fact that intramolecular association no longer dominates for large $\varepsilon^*$. The general trend, for the studied chain lengths, is that increasing chain length increases intermolecular association and decreases intramolecular association. The reason for this is easily seen in $P_{form}(m)$ in table 1. As chain length is increased, the probability that the chain will be in a conformation such that intramolecular association can occur decreases. Since fewer chains are associated in ring states more association sites are available for intermolecular association. A major exception is for the case $m = 3$, which shows the strongest degree of intermolecular association and no intramolecular association; this is due to the fact that it is impossible for this short molecule to intramolecularly associate into a ring.

Now we will study how the presence of a hard wall affects the bonding fractions. Here we will consider the ratios $R_{chain}(z) = \chi_{chain}(z)/\chi_{chain}^{bulk}$ and $R_{ring}(z) = \chi_{ring}(z)/\chi_{ring}^{bulk}$, which show how the hard wall affects the bulk association. Figure 8 gives these ratios for chains of lengths $m$ = 4 – 7, at association energies $\varepsilon^* = 6$, 10, 12. We begin our discussion with the low energy case $\varepsilon^* = 6$. Both ratios $R_{ring}$ and $R_{chain}$ are depleted at wall contact ($z = 0$). This shows that the wall hinders both intra and intermolecular association, with intermolecular association being more hindered than intramolecular association, $R_{chain}(0) < R_{ring}(0)$. In each case, increasing chain length further depletes association at wall contact. As we move away from the wall $R_{chain}$ increases, but at no point becomes significantly greater than 1, and the chain length dependence of $R_{chain}$ does not change. In contrast, $R_{ring}$ becomes enhanced as we move away from the wall with a maximum near $z = \sigma$, and the chain length dependence of $R_{ring}$ inverts (at wall contact increasing $m$ decreases $R_{ring}$ while the opposite is true at z ~ 1.5σ). The depletion of $R_{chain}$ is easy to understand. There are less available configurations that two chains can take near the wall where association can occur, resulting in an entropic penalty, the longer the chain the larger penalty. Also, at this packing fraction the density of chains becomes depleted at the wall further hindering intermolecular association. Moving away from the wall lessens these penalties resulting in an increase in $R_{chain}$. The behavior of $R_{ring}$ can be understood in terms of the single chain probability $P_{form}(z, m)$, which is the generalization of the probability $P_{form}(m)$ given by Eq. (36) to inhomogeneous systems. This is simply the probability that if segment 1 is located at position z near a hard wall that segment $m$ is positioned and oriented such that association can occur, that is

$$P_{form}(z, m) = \frac{\int d\Omega_1 d\Omega_m d\vec{r}_1 \ldots d\vec{r}_m \delta(z_1 - z) U(\Omega_1, \Omega_m) P_{chain}(\vec{r}_1 \ldots \vec{r}_m) \xi(r_{1m}) \prod_{k=1}^{m} H(z_k)}{\Omega_1 \Omega_m \int d\vec{r}_1 \ldots d\vec{r}_m \delta(z_1 - z) P_{chain}(\vec{r}_1 \ldots \vec{r}_m) \prod_{k=1}^{m} H(z_k)} \qquad (43)$$

Where $H(x)$ is the Heaviside step function. We evaluated Eq. (43) using monte carlo integration, the results can be found in Fig. 9. We note that the ratio $R_{form}(z, m) = P_{form}(z, m)/P_{form}(m)$ is depleted at wall contact showing there is a lower probability the chain is in a ring conformation as compared to the bulk, increasing chain length results in further depletion of $R_{form}$. Moving away from the wall $R_{form}$ goes through a maximum and then the chain length dependence inverts, such that increasing $m$ results in an increase in $R_{form}$ at a given position $z$. The inversion of the chain length dependence is explained as follows; near the wall the number of accessible chain and ring states are smaller than the bulk, with the number of ring states being the most depleted, giving a $R_{form} < 1$. However, as we move away from the wall, at some point the number of accessible chain states becomes more depleted than the number of ring states. This is the region where $R_{form} > 1$. The longer the chain the stronger and more long ranged this effect, resulting in an inversion of the chain length dependence of $R_{form}$. Comparing the ratio $R_{form}$ in Fig. 9 and $R_{ring}$ at $\varepsilon^* = 6$ in Fig. 8, it is clear that the positional dependence of intramolecular association is dominated by the probability that the two chain ends are positioned and oriented such that intramolecular association can occur.

As association energy is increased to $\varepsilon^* = 10$ (center panel Fig. 8) and $\varepsilon^* = 12$ (right panel Fig. 8), the ratio contact values $R_{ring}(0)$ become enhanced and the chain length dependence begins to change such that increasing $m$ increases $R_{ring}(0)$; simultaneously the chain length dependence of $R_{chain}$ inverts to where increasing $m$ results in an increase in $R_{chain}$. This is exactly the opposite behavior observed at $\varepsilon^* = 6$. This change in behavior is the result of the competition between intra and intermolecular association and is not observed in systems which only intra or intermolecularly associate (not both); this can be seen in Fig. 10 which shows the ratios $R_{ring}$ for a system with only intramolecular association and $R_{chain}$ for a system only with intermolecular association, each at $\varepsilon^* = 12$. As can be seen, the ratios in Fig. 10 mimic the low $\varepsilon^* = 6$ case from Fig. 8 and do not show the change in chain length dependence of $R_{chain}$ and $R_{ring}(0)$. Also the contact values $R_{ring}(0)$ at no point become enhanced as seen in Fig. 8.

The contact values $R_{ring}(0)$ become enhanced at high $\varepsilon^*$ when both intra and intermolecular association are possible (Fig. 8) due to the fact that the wall inhibits intermolecular association to a greater degree than intramolecular association (contact values Fig. 10). At these high association energies (low temperatures) association is desired and the easiest way to accomplish this near wall contact is intramolecular association. For this reason, $R_{ring}(0)$ becomes enhanced at high $\varepsilon^*$ and $R_{chain}(0)$ remains depleted.

The change in chain length dependence of $R_{chain}$ is also the result of intramolecular association being favored near wall contact. Consider the bottom right panel of Fig. 8 for $R_{chain}$ at $\varepsilon^* = 12$, when both intra and intermolecular association are possible, and the bottom panel of Fig. 10, when only intermolecular association is possible. The addition of intramolecular association causes $R_{chain}$ to decrease. This is due to the fact that intramolecular association is favored in the inhomogeneous region; the stronger the tendency to intramolecular associate into rings, the more pronounced the decrease in $R_{chain}$. Since shorter molecules show the strongest degree of intramolecular association (for $m > 3$), we see that increasing $m$ results in an increase in $R_{chain}$. This is the exact opposite trend observed when intramolecular association is not allowed.

Finally we plot the contact values $R_{chain}(0)$ and $R_{ring}(0)$ as a function of $\varepsilon^*$, at a bulk packing fraction of $\eta = 0.1$ in Fig. 11. In addition, Fig. 12 gives the ratio $R_{total}(0) = \chi(0)/\chi^{bulk}$, showing that even at wall contact nearly all end segments form association bonds at large enough $\varepsilon^*$. We see that $R_{chain}(0)$ for $m = 3$ goes to unity as $\varepsilon^* \to \infty$, this results from the fact that at large enough $\varepsilon^*$ the enthalpic benefit of association far outweighs the entropic penalty imposed by the presence of the wall, and since only intermolecular association is possible there is no competition with ring formation.

For $m = 4$ both $R_{chain}(0)$ and $R_{ring}(0)$ initially increase with $\varepsilon^*$, then both quantities reach maximums, and then begin to decrease as a function of $\varepsilon^*$ with $R_{ring}(0) \to 1$ as $\varepsilon^* \to \infty$. The maximum in $R_{ring}(0)$ results from the fact that as $\varepsilon^*$ is increased the competition between intra and intermolecular association results in an enhancement of $R_{ring}(0)$. However, at high enough $\varepsilon^*$ both the fraction of end segments bonded (either intermolecularly or intramolecularly) $\chi$ and the fraction bonded intramolecularly $\chi_{ring}$ approach unity throughout the computational domain resulting in the limit $R_{ring}(0) \to 1$ as $\varepsilon^* \to \infty$. The behavior of $R_{chain}(0)$ is explained as follows:

as $\varepsilon^*$ is increased, $R_{chain}(0)$ begins to increase in the same manner as for $m = 3$; again, this increase is the result of the higher enthalpic benefit of association at larger $\varepsilon^*$. However, intramolecular association becomes dominant for $m = 4$, and intramolecular association is favored at wall contact over intermolecular association; this results in the maximum in $R_{chain}(0)$ and the eventual decrease to a value of $R_{chain}(0) \sim 0.37$ for large $\varepsilon^*$.

Increasing the chain length to $m = 5 - 7$ the maximums in $R_{ring}(0)$ and $R_{chain}(0)$ disappear. Also the ratio $R_{ring}(0)$ approaches a limiting value less than 1. The change in behavior results from the fact that the system does not become completely dominated by rings for $\varepsilon^* \to \infty$ as in the case $m = 4$. Figure 11 nicely demonstrates the reversal of the chain length dependence of $R_{chain}(0)$ and $R_{ring}(0)$ as $\varepsilon^*$ is increased. Again, this reversal is the result of the competition between inter and intramolecular association and is not seen in systems which only inter or intramoleculary associate.

## VI: Conclusions

We have developed the first MCDFT for associating chain molecules. The theory is applicable to chains with an association site on each end of the chain. Both intra and intermolecular association are taken into account. The theory was tested against monte carlo simulation data for a 4-mer in a hard planar slit pore. For an average pore packing fraction of $\eta_{av} = 0.1$ the theory was found to be in excellent agreement with simulation data for the density profiles, fraction of end segments associated $\chi$ and the fraction of end segments associated intramolecularly $\chi_{ring}$. Increasing the average pore packing fraction to $\eta_{av} = 0.3$ the theory was in excellent agreement with simulation for the density profiles and $\chi$; however, the theory overpredicted the fractions $\chi_{ring}$. The theory was then applied to study how the presence of a hard wall changed the bonding fractions in relation to their bulk value as a function of chain length and association energy. It was shown that the competition between inter and intramolecular association enhances intramolecular association near wall contact and inverts the chain length dependence of the fraction bonded intermolecularly in the inhomogeneous region. This was the first study to determine the effect of chain length and confinement on intra and intermolecularly associating chain molecules. This type of system can provide insight in into the behavior of glycol ethers and telechelic polymers near solid surfaces.

## Acknowledgements

The financial support for this work was provided by the Robert A. Welch Foundation (Grant No. C-1241). A.J.G.C acknowledges support from Tecnológico de Monterrey (Grant No. CAT-125).

## Table Captions:

**Table 1:** Numerical calculations of $P_{form}(m)$

## Figure Captions:

**Figure 1:** Diagram of model associating chain molecule

**Figure 2:** Comparison of theoretical density profiles (dashed line - middle segment, solid line - end segment) to NVT monte carlo simulations (circles – end segment, diamonds – middle segment) at two average packing fractions and two association energies

**Figure 3:** Fraction of end segments bonded intramolecularly for $\eta_{av} = 0.1$ (top) and $\eta_{av} = 0.3$ (bottom). Symbols give simulation results and curves are MCDFT predictions

**Figure 4:** The quantity $\Delta^{intra} / f_{AB}$ for a 4-mer as a function of bulk packing fraction. Curve gives theoretical predictions (Eq. 41) and symbols are the monte carlo simulation results of Ghonasgi and Chapman[17]

**Figure 5:** Bonding fractions $\chi$ (top) and $\chi_{ring}$ (bottom) for $\eta_{av} = 0.1$. Symbols give simulation results and curves are MCDFT predictions

**Figure 6:** Same as Fig. 5 with $\eta_{av} = 0.3$

**Figure 7:** Fraction of end segments bonded intramolecularly $\chi_{ring}$ (top) and fraction of end segments bonded intermolecularly $\chi_{chain}$ (bottom) in a bulk system at a packing fraction of $\eta_b = 0.1$ for chain lengths $m$ = 3 - 7

**Figure 8:** Ratios $R_{ring}(z) = \chi_{ring}(z) / \chi_{ring}^{bulk}$ and $R_{chain}(z) = \chi_{chain}(z) / \chi_{chain}^{bulk}$ for a fluid near a hard wall with a bulk packing fraction $\eta_b = 0.1$

**Figure 9:** The probability $P_{form}(z, m)$ that if segment 1 is located at position z in the pore that segment $m$ is positioned and oriented such that association can occur. The probability is scaled by the bulk probability $P_{form}(m)$

**Figure 10:** Ratios $R_{ring}$ for $\varepsilon* = 12$ and $\eta_b = 0.1$ when only intramolecular association is allowed (top) and $R_{chain}$ when only intermolecular association allowed (bottom)

**Figure 11:** Contact values $R_{ring}(0) = \chi_{ring}(0) / \chi_{ring}^{bulk}$ and $R_{chain}(0) = \chi_{chain}(0) / \chi_{chain}^{bulk}$ for a fluid near a hard wall with a bulk packing fraction $\eta_b = 0.1$

**Figure 12:** Contact values $R_{total}(0) = \chi(0) / \chi^{bulk}$ for a bulk packing fraction $\eta_b = 0.1$

**Table 1:**

| $m$ | $P_{form}(m) \, x10^5$ |
|---|---|
| 3 | 0 |
| 4 | 20.8 |
| 5 | 7.3 |
| 6 | 3.0 |
| 7 | 1.7 |
| 8 | 1.1 |

**Figure 1:**

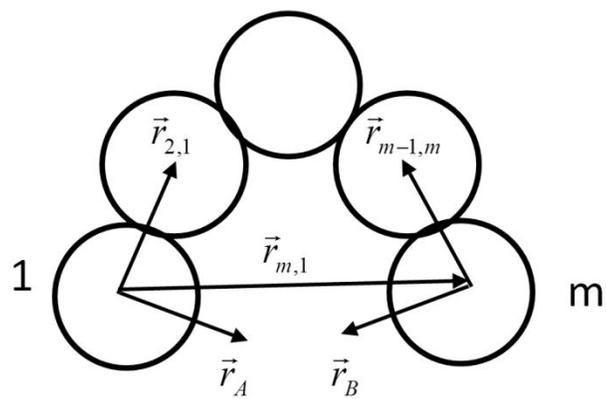

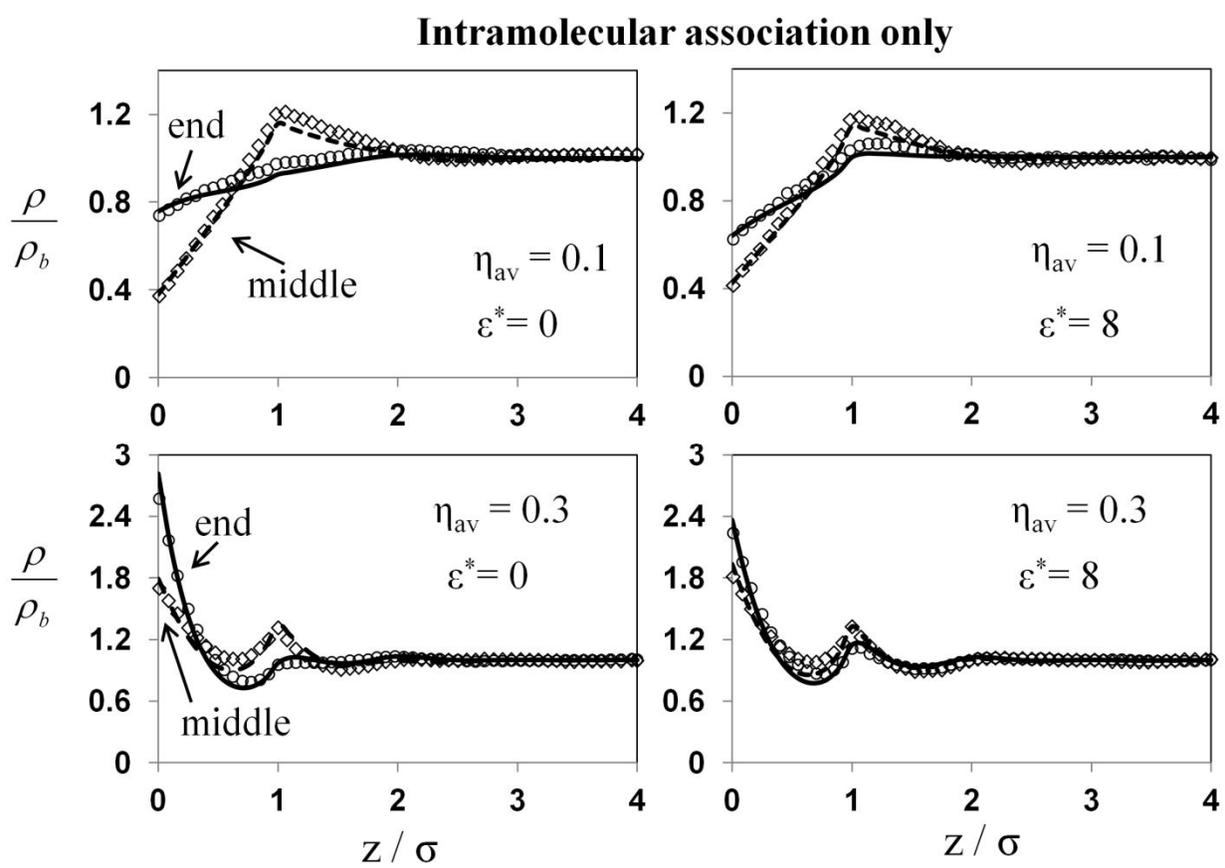

**Figure 3:**

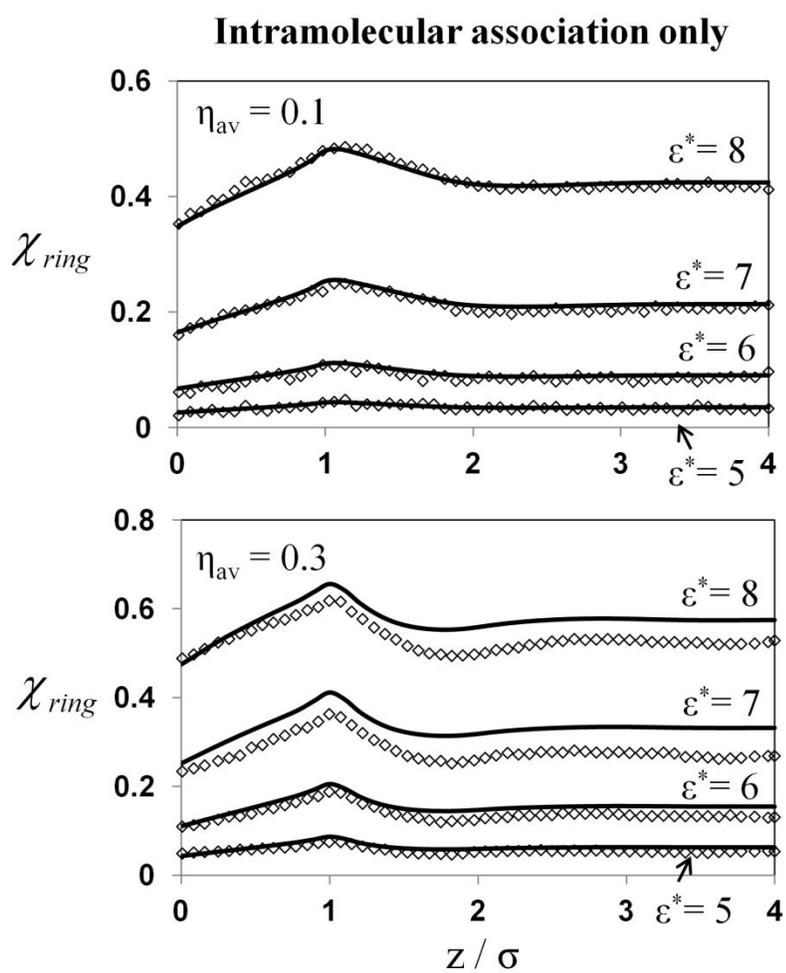

**Figure 4:**

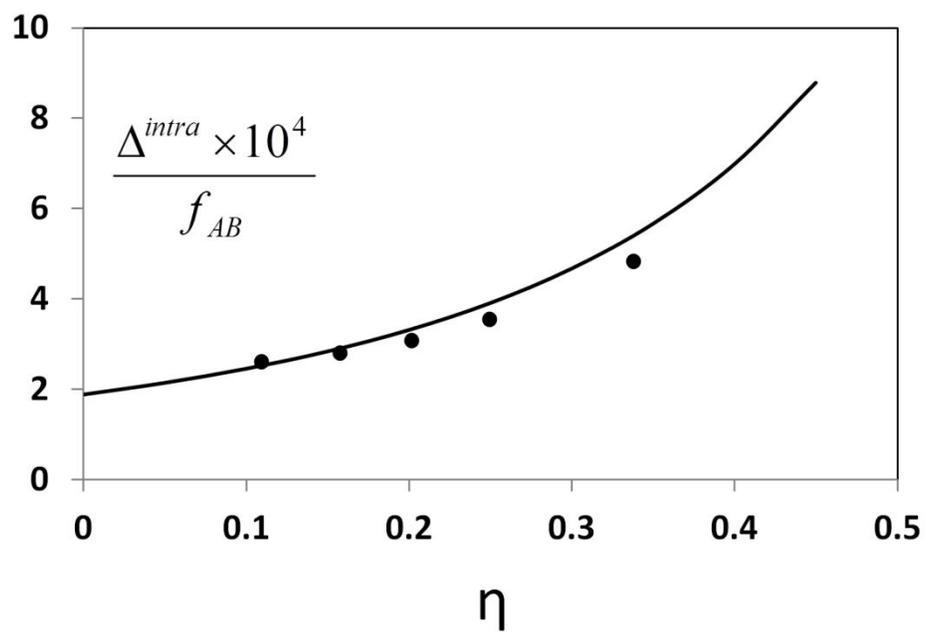

**Figure 5:**

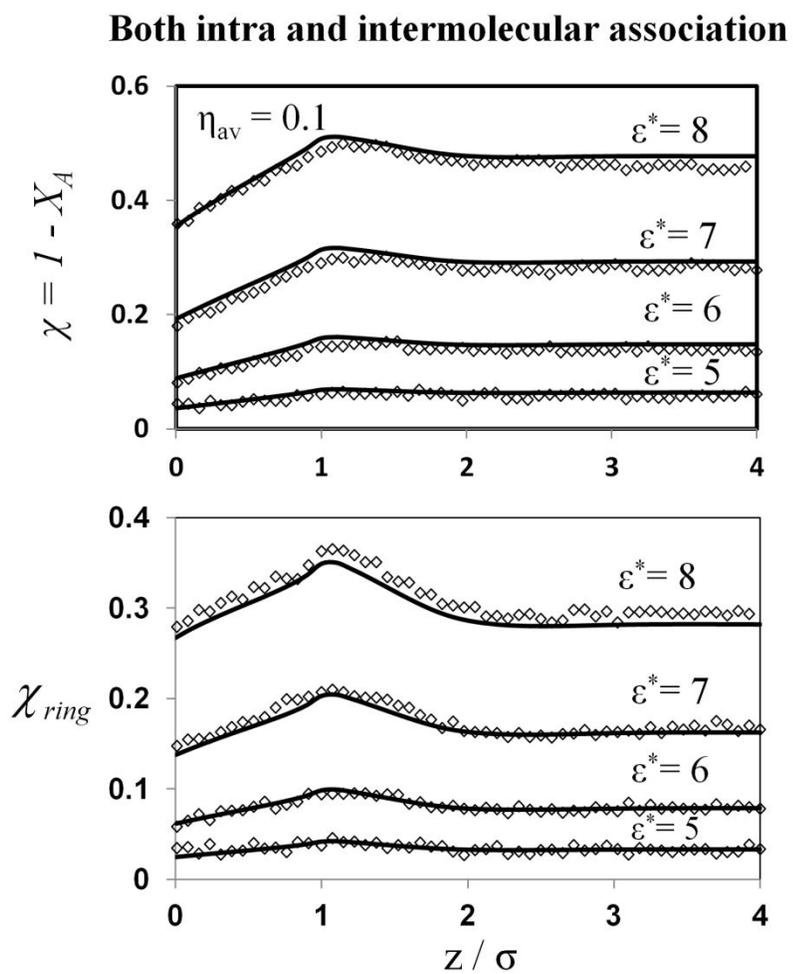

Both intra and intermolecular association

**Figure 6:**

**Both intra and intermolecular association**

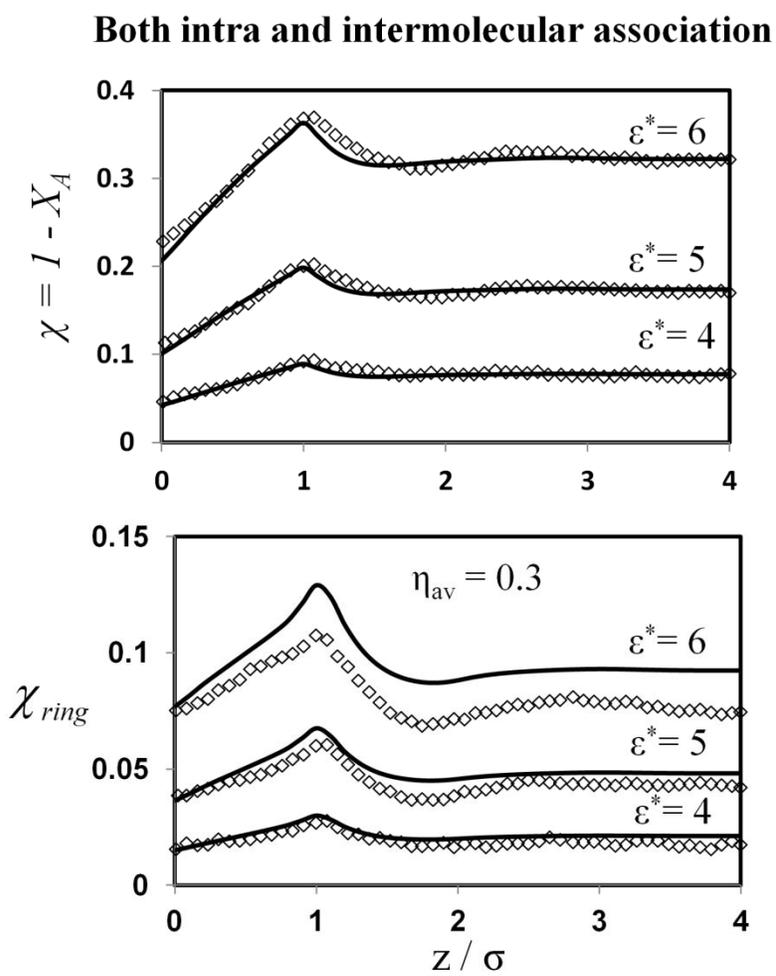

**Figure 7:**

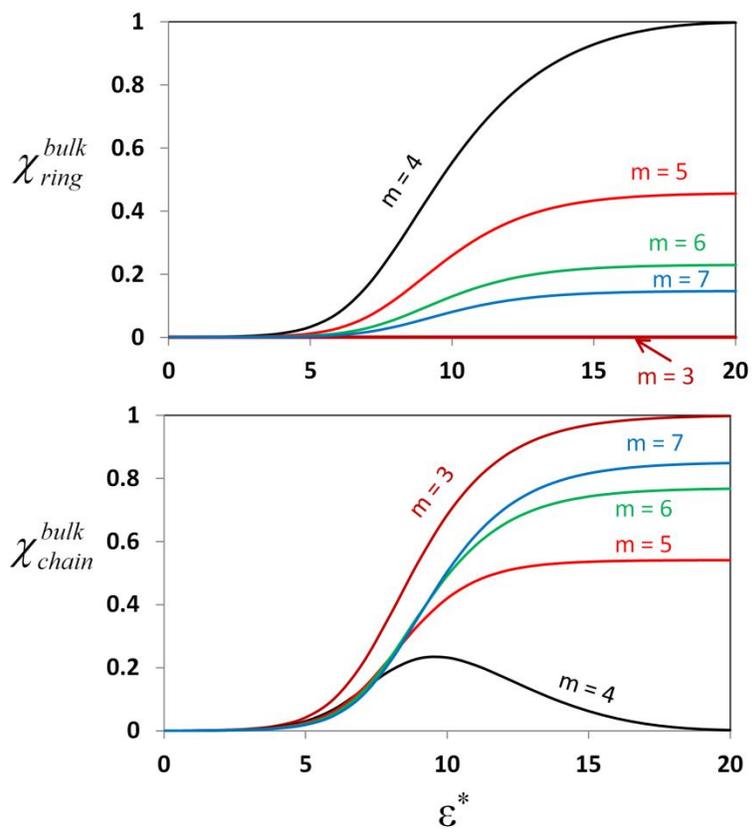

**Figure 8:**

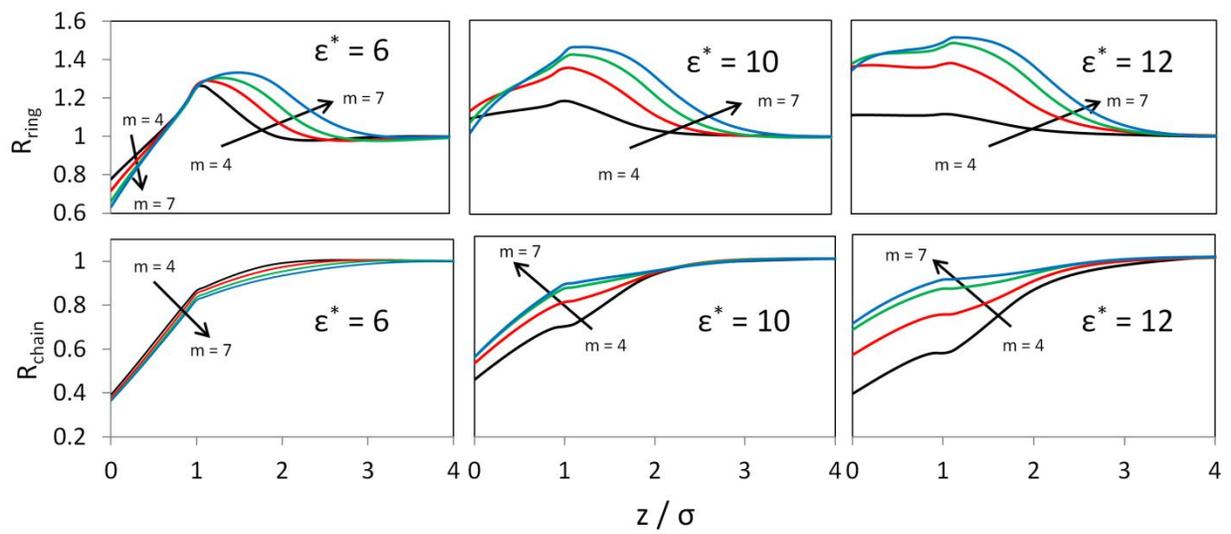

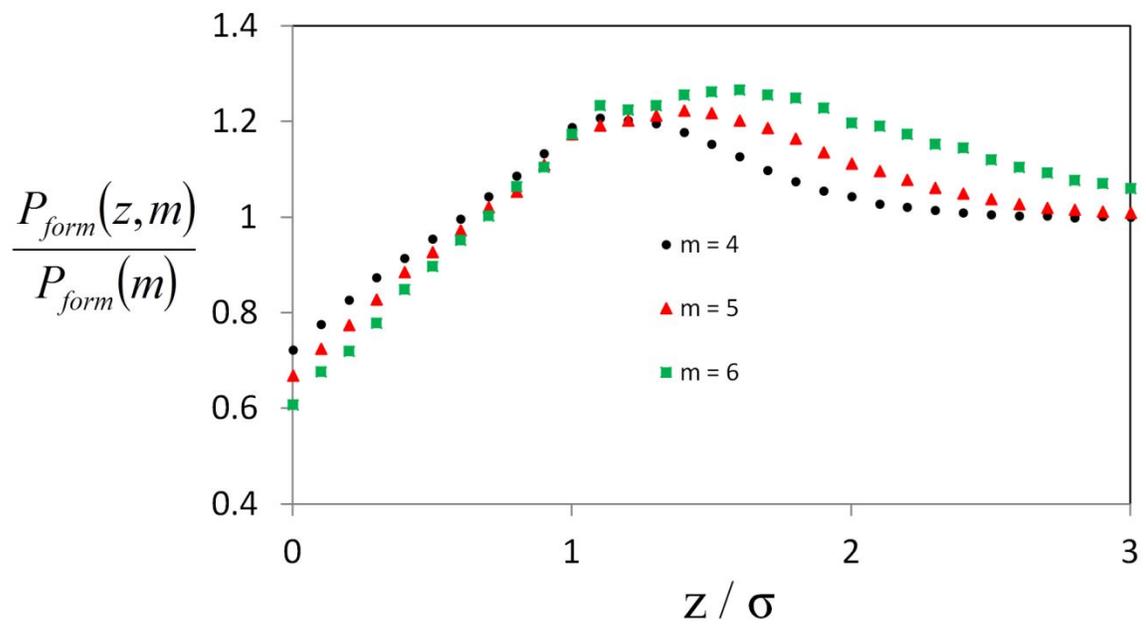

**Figure 10:**

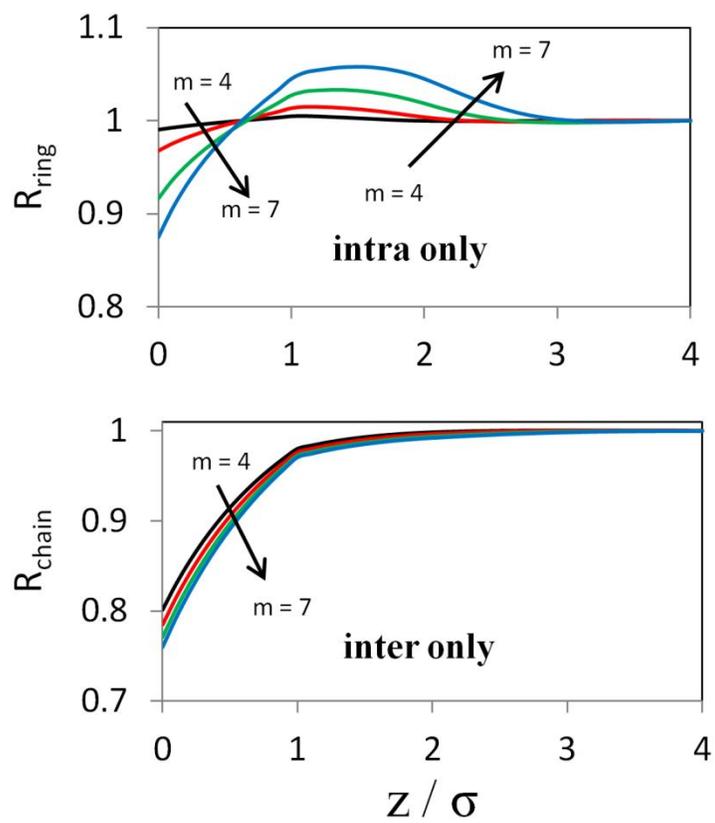

**Figure 11:**

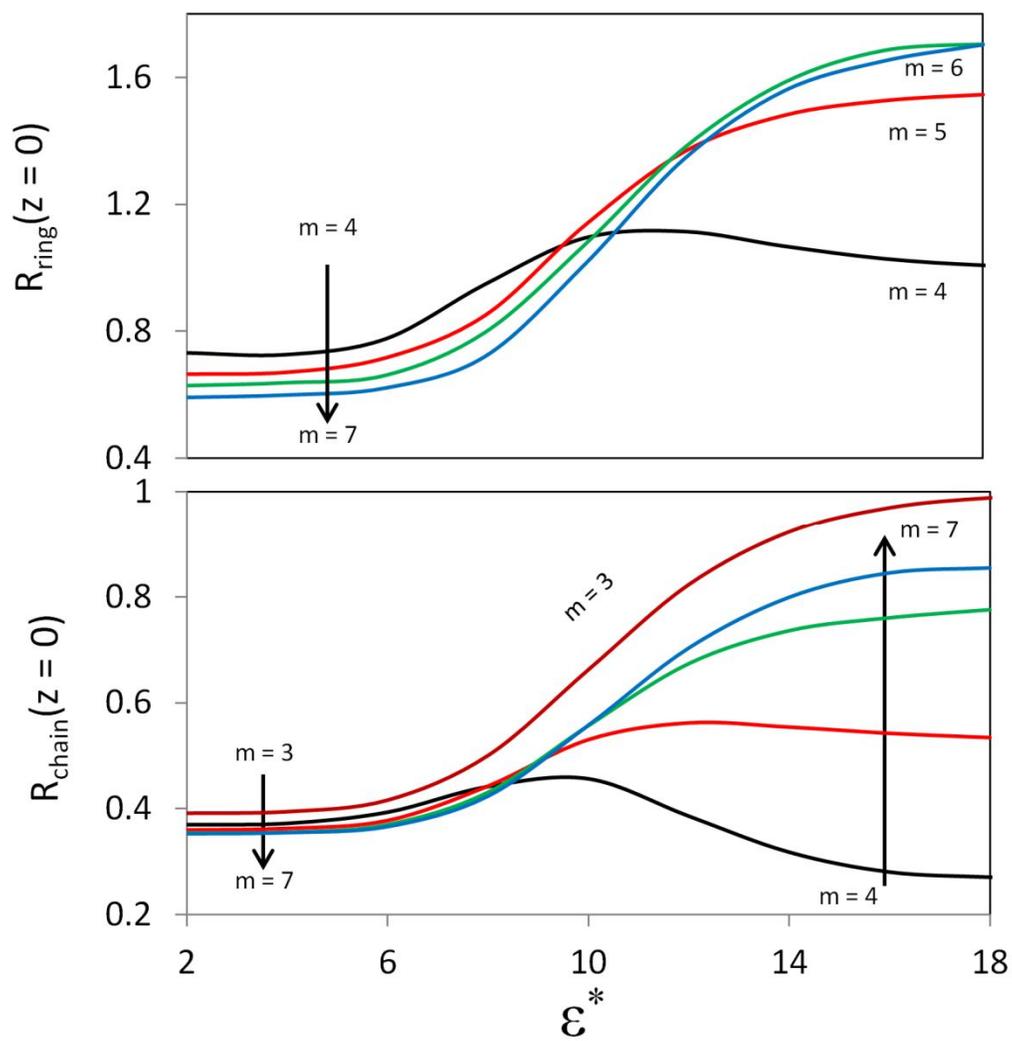

**Figure 12:**

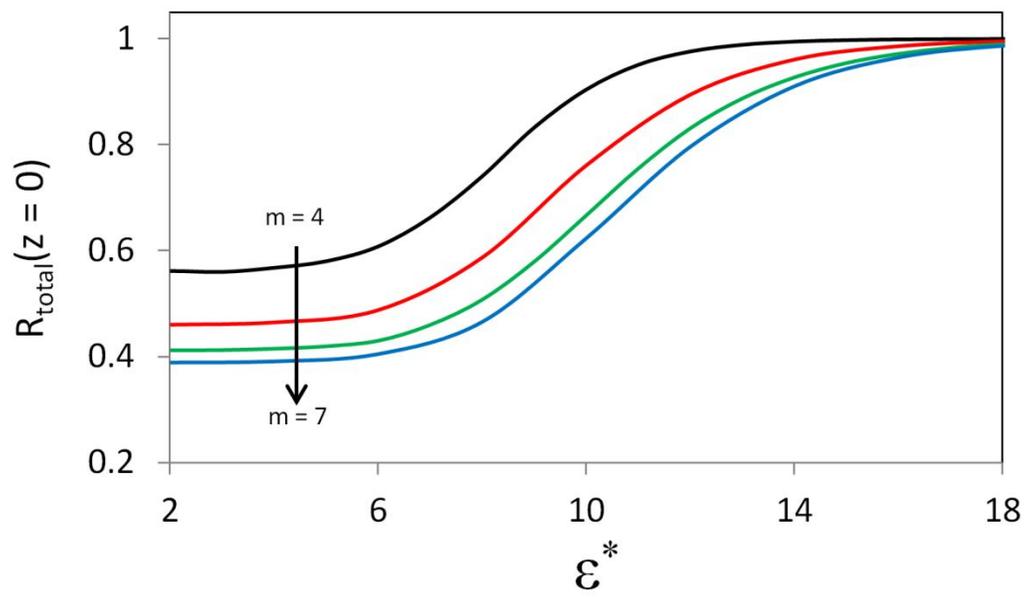